\pdfoutput=1
\documentclass[
nofootinbib,amsfonts,amssymb,amsmath
, twocolumn
]{revtex4-2}
\usepackage{amsmath,bm,mathtools,physics}
\usepackage[utf8]{inputenc}
\usepackage[oldstyle]{libertine}
\usepackage{libertinust1math}
\numberwithin{equation}{section}
\renewcommand\theequation{\arabic{section}.\arabic{equation}}
\usepackage{fontawesome}

\usepackage{graphicx}
\usepackage{comment}
\graphicspath{ {/} }

\newcommand{\equref}[1]{eqn.~(\ref{#1})}
\allowdisplaybreaks

\def\dotl[#1,#2]{\left\langle #1,\, #2 \right\rangle}
\def\dotlb[#1,#2]{\left\langle #1,\, #2 \right\rangle}
\def\dotlm[#1,#2]{\left[ #1,\, #2 \right]}
\def\dotp[#1,#2]{(\vect{#1} \cdot\vect{#2})}
\def\lb{\bar{\lambda}}
\def\be#1\ee{\begin{equation}\begin{aligned}#1\end{aligned}\end{equation}}
\makeatletter
\DeclareRobustCommand\bea{\@ifnextchar[{\@@bea}{\@bea}}
\def\@@bea[#1]#2\eea{\begin{subequations}\begin{align}#2\end{align}\label{#1}\end{subequations}}
\def\@bea#1\eea{\begin{subequations}\begin{align}#1\end{align}\end{subequations}}
\makeatother
\newcommand \nn {\nonumber}
\def\<#1\>{\expval{#1}}
\newcommand   \x   {\times}
\newcommand   \e  {\epsilon}
\newcommand   \s  {\sigma}
\renewcommand \a  {\alpha}
\newcommand   \de {\delta}
\newcommand   \w  {\omega}

\newcommand   \cV {\mathcal{V}}
\newcommand   \cB {\mathcal{B}}
\newcommand   \cA {\mathcal{A}}
\newcommand   \cC {\mathcal{C}}
\newcommand   \cI {\mathcal{I}}
\newcommand   \cN {\mathcal{N}}
\newcommand   \cO {\mathcal{O}}
\newcommand   \cD {\mathcal{D}}
\newcommand   \SO    {\mathrm{SO}}
\newcommand   \SL    {\mathrm{SL}}
\newcommand   \R  {\mathbb{R}}
\newcommand   \C  {\mathbb{C}}
\newcommand   \half{\frac 1 2}
\def\bk{\bm{k}}
\def\bki{\bm{k}_{\text{int}}}
\def\ki{k_{\text{int}}}

\def\bep{\bm{\e}}
\def\lb{\bar{\lambda}}

\def\lib{\mathfrak{\bar i}}
\def\li{\mathfrak{i}}

\usepackage{color}
\definecolor{darkblue}{rgb}{0.1,0.1,.7}
\usepackage[colorlinks, linkcolor=darkblue, citecolor=darkblue, urlcolor=darkblue, linktocpage]{hyperref} 
\begin{document}
\title{All plus four point (A)dS graviton function using generalized on-shell recursion relation}

\def\andname{}
\author{Soner Albayrak$^{\text{\faDribbble}}$}
\author{Savan Kharel$^{\text{\faDotCircleO}}$}
\affiliation{$^{\text{\faDribbble}}$Institute of Physics, University of Amsterdam, Amsterdam, 1098 XH, The Netherlands}
\affiliation{$^{\text{\faDotCircleO}}$Department of Physics, University of Chicago, Chicago, IL 60637, USA}

\begin{abstract}
This paper presents a calculation of the four gravitons amplitude in (Anti)-de Sitter space, focusing specifically on external gravitons with positive helicity. To achieve this, we employ a generalized recursion method that involves complexifying all external momentum of the graviton function, which results in the factorization of AdS graviton amplitudes and eliminates the need for Feynman-Witten diagrams. Our calculations were conducted in three boundary dimensions, with a particular emphasis on exploring cosmology and aiding the cosmological bootstrap program. To compute the expression, we utilized the three-dimensional spinor helicity formalism. The final expression was obtained by summing over residues of physical poles, and we present both symbolic and numerical results. Additionally, we discuss the advantages and limitations of this approach, and highlight potential opportunities for future research.
\end{abstract}
\date{\today}
\maketitle

\section{Introduction}
The surprising simplicity of scattering amplitudes in flat space has provided a beautiful conundrum: why is the scattering amplitudes of gravity simple, given the non-linear nature of general relativity? Attempts to understand this underlying elegance has driven important advancements in theoretical physics in recent years \cite{Elvang:2013cua}. One of the most effective tools for studying the simplicity of these observables in flat space is the on-shell recursion relations, which express the scattering amplitudes of higher point amplitudes as a sum of products of lower-point amplitudes \cite{Britto:2005fq}.

Alongside the advances in Minkowski space, there has been a growing emphasis on the study of holographic observables. Holography finds concrete appearance in asymptotically Anti-de Sitter (AdS) spacetimes, where AdS scattering amplitudes are equivalent to the correlation functions of a conformal field theory (CFT) on the boundary \cite{Maldacena:1997re}. Many innovative techniques such as Mellin space, conformal bootstrap, and harmonic analysis in AdS are used in the computations of these correlators \cite{Penedones:2010ue,Fitzpatrick:2011ia,Rastelli:2016nze, Heemskerk:2009pn,Costa:2014kfa,Sleight:2017fpc, Fitzpatrick:2011dm, Costa:2014kfa, Paulos:2011ie, Kharel:2013mka}. This work will build upon progresses made in (A)dS momentum space \cite{Bzowski:2013sza, Bzowski:2015pba, Bzowski:2018fql, Bzowski:2015pba, Bzowski:2019kwd, Bzowski:2020kfw, Isono:2018rrb, Isono:2019wex, Coriano:2013jba, Coriano:2018bbe, Anand:2019lkt, Farrow:2018yni, Nagaraj:2019zmk, Nagaraj:2020sji, Jain:2020rmw, Meltzer:2020qbr, Albayrak:2018tam, Albayrak:2019asr, Albayrak:2019yve, Albayrak:2020isk, Albayrak:2020bso, Bzowski:2022rlz, Caloro:2022zuy, Meltzer:2021zin}. However, we use an approach which has been largely under-used in curved spacetime. Instead of relying on perturbative Feynman-Witten diagrams, we will use recursion relations to compute our expression, revisiting an attractive perspective on the topic that was started in these seminal papers \cite{Raju:2012zr, Raju:2012zs}.

A compelling reason to investigate these correlators in momentum space is due to its close relationship to the \emph{wave function of the universe} \cite{Maldacena:2002vr,Ghosh:2014kba}. Therefore, the cosmological bootstrap \cite{Arkani-Hamed:2018kmz,Baumann:2019oyu,Baumann:2020dch,Sleight:2019hfp,Sleight:2020obc} and cosmological collider physics programs \cite{Arkani-Hamed:2015bza, Wang:2022eop, Niu:2022fki, Niu:2022quw, Xianyu:2022jwk, Qin:2022lva, Reece:2022soh, Reece:2022soh, Heckelbacher:2022hbq, Pinol:2021aun, Lu:2021wxu, DiPietro:2021sjt, Kumar:2019ebj, Hook:2019vcn, Hook:2019zxa, Alexander:2019vtb, Aoki:2023tjm, Qin:2023ejc, Pimentel:2022fsc, Tong:2022cdz, Baumann:2019ztm} have been putting an organized effort to unravel the properties of cosmologically relevant correlators. While the language of AdS computations looks different, the techniques here translate naturally to cosmology as both AdS and cosmological correlators exhibit a total energy singularity when sum of the norms of the momenta are analytically continued to zero. The behavior of the correlator around this singularity is closely linked to the scattering amplitude for the same process in flat space, meaning that holographic and cosmological correlators contain within them valuable physics of the flat space scattering amplitudes.

In curved spaces, on-shell recursion relations require various modifications,\footnote{Firstly, it is not a priori given that a correlator in general curved spaces is a meromorphic function, hence the usual contour-deformation arguments of BCFW needs to be altered to take into account the other singularities. Secondly, the absence of the conservation of momentum along the bulk direction complicates the deformation. Luckily, flat-space-like on-shell recursion relations do exist in (A)dS, and one can work with a simple deformation by sticking to momentum space in the Poincar\'e patch as we detail in the body.} which is especially true for the three-dimensional boundaries that are most relevant to cosmology. In three dimensions, the standard two-line shift used in BCFW recursion relations is not possible, so a generalization to an $n$-line deformation is necessary \cite{Raju:2012zs}. However, this significantly increases the number of partitions to consider, which reduces efficiency. Moreover, the complexity of the computation is further increased by the fact that the deformation parameter satisfies a quadratic equation, as the boundary momenta are not null. These challenges have hindered the use of  {\itshape RBCFW} (Raju's extension of BCFW in AdS), developed by Raju over a decade ago. In our work, we aim to clarify these challenges and demonstrate the areas where (A)dS on-shell methods are most effective.

Our paper aims to calculate the four point graviton amplitude with all plus external helicities using RBCFW and spinor helicity techniques. There have been other recent approaches to this problem where the authors compute late time four point function and the related quartic coefficient by direct computation and by other approaches  \cite{Bonifacio:2022vwa}. Similarly, authors of this paper have computed up to five point exchange diagrams in AdS in \cite{Albayrak:2019yve}.  However, we will proceed with RBCFW instead, and compute four point all plus helicity graviton amplitude. Besides being previously uncomputed, all plus helicity has also been historically important in the flat space scattering amplitude program as a playground to discover structures about more complicated amplitudes. Interestingly, the four point flat space all plus helicity amplitude vanishes at tree-level; but for pure Yang-Mills and gravity, the loop level counterpart is non-zero for non-supersymmetric cases. These amplitudes have been used as toy examples to study non-supersymmetric amplitudes in flat space. Nevertheless, the task to compute this function in (A)dS for graviton is rather formidable. It's important to remember that even in flat space, computing the four point graviton amplitude is a challenging task that was only made easier with the development of on-shell recursion techniques.  However, in principle, using RBCFW is a straightforward approach to tackle these calculations. Moreover, recursive methods can be employed to compute higher point amplitudes without having to worry about the infinite series of interaction vertices that arise from the expansion of the Einstein-Hilbert action. Therefore, we believe that this is an intriguing approach that demands more attention.

The paper is structured as follows. In \S~\ref{sec:preliminaries}, we briefly review the momentum space perturbation theory and on-shell methods in AdS via providing the necessary ingredients for the calculations. In \S~\ref{sec:main}, we compute the all plus helicity four point graviton amplitude and analyze it from different analytical and numerical perspectives. We then conclude in \S~\ref{sec:conclusion} by discussing the efficacy of RBCFW computations as to the weaknesses and strengths of this technology. In particular, we detail in what kind of computations we believe that this method would shine, and then close with an outlook. We collect the technical details in the appendix, along with a review of the spinor helicty formalism in AdS.\footnote{The attached Mathematica files contain the less presentable data, such as the symbolic expression for the full amplitude and the data points for Figure~\ref{figure}.}

\section{Preliminaries}
\label{sec:preliminaries}
\subsection{Momentum space perturbation theory in AdS}
We start by setting our conventions and notations. We work in Poincar\'e coordinates with the metric \mbox{$ds^2=z^{-2}\left(dz^2+\eta_{ij}dx^idx^j\right)$}: $z$ is the radial coordinate, $x_i$ approach the boundary as $z\rightarrow 0$, and $\eta_{ij}$ is the boundary metric in mostly positive signature. As we preserve the manifest translation symmetry at the boundary, we will make use of it by going to the boundary Fourier domain and working with the coordinates $\{z,\bk_i\}$, following the treatment of the related papers in the literature \cite{Raju:2010by,Raju:2011mp,Raju:2012zs,Raju:2012zr,Albayrak:2018tam,Albayrak:2019asr,Albayrak:2019yve,Albayrak:2020isk,Albayrak:2020bso,Albayrak:2020fyp}.

The perturbative treatment of gravity in AdS is relatively complicated albeit straightforward:\footnote{The quadratic part of the gravity action can be used to obtain the bulk to bulk propagator, from which one can obtain the bulk to boundary propagator by taking one of the vertices to the boundary.} we will simplify the expressions by sticking to the \emph{axial gauge}, i.e. $h_{\mu0}=0$. In this gauge, the bulk-to-boundary propagator reads as
\begin{subequations}
	\label{eq: gravity propagators}
	\begin{equation}
		h_{ij}(\bm{k},z)=\bm{\epsilon}_i\bm{\epsilon}_j\sqrt{\frac{2}{\pi}}z^{-2}(k z)^{\frac{d}{2}} K_{\frac{d}{2}}(k z)\;,
	\end{equation}
	whereas the bulk to bulk propagator is
	\be 
	\label{eq: graviton propagator}
	\mathcal{G}_{ab,cd}(\bm{k}; z, z')={}&
	\frac{i(zz')^{\frac{d}{2}-2}}{2}\int\limits_{0}^{\infty} dp 
	J_{\frac{d}{2}}(p z)J_{\frac{d}{2}}(p z')
	\\\hspace*{-.05in}
	&\hspace*{-6em}{}\x\frac{p \left(H^{(p,\bm{k})}_{ac}H^{(p,\bm{k})}_{bd}+H^{(p,\bm{k})}_{ad}H^{(p,\bm{k})}_{bc}-\frac{2}{d-1}H^{(p,\bm{k})}_{ab}H^{(p,\bm{k})}_{cd}\right)}{k^2+p^2-i \epsilon}
	\ee 
\end{subequations}

Here, $k_i$ denote the norm of $\bk_i$, $\bm{\e}_i$ is the polarization vector, and the tensor $H$ is defined as \mbox{$H_{ij}^{(p,\bm{k})}\coloneqq{}-i\left(\eta_{ij}+{\bm{k}_i\bm{k}_j}/p^2\right)$}. Finally, the vertex factor for the graviton cubic self-interaction is\footnote{Note the contravariant form for the vertex factor (hence $z^8$).}
\begin{multline}
	\label{eq: graviton ingredients}
	\cV^{ijklmn}_{\bm{k}_1, \bm{k}_2, \bm{k}_3}=z^8\bigg(
	\frac{(\bm{k}_2)^i(\bm{k}_3)^j\eta^{km}\eta^{ln}}{4}
	\\-\frac{(\bm{k}_2)^i(\bm{k}_3)^k\eta^{jm}\eta^{ln}}{2}
	\bigg)+\text{ permutations}
\end{multline}
We refer the reader to \cite{Raju:2011mp} for further details on the derivation of these ingredients.

The computation of Witten diagrams in this formalism is quite straightforward: we take the contracted product of necessary ingredients and then integrate over bulk radius $z$ of internal vertices; for instance, the three point graviton amplitude has the expression
\be 
\cA^{\text{gr}}_3=\int\limits_0^\infty\frac{dz}{z^{d+1}}h_{ij}(\bk_1,z)h_{kl}(\bk_2,z)h_{mn}(\bk_3,z)\cV^{ijklmn}_{\bk_1,\bk_2,\bk_3}
\ee 

The Witten diagrams computed along these lines correspond to the vacuum correlators with all sources in the same Poincar\'e patch. In contrast, general correlation functions in global AdS may have sources in multiple patches; when viewed from one patch, insertions in the other patches are invisible and they simply amount to creating boundary conditions on the past and future horizons of the chosen patch \cite{Raju:2011mp}. Such objects are called \emph{transition amplitudes}, and we can derive them from the usual Witten diagrams by a simple replacement of bulk to boundary propagators (corresponding to the insertions in the other patches) with normalizable modes: in practice, this is equivalent to the replacement of the modified Bessel function of the second kind ($K$) with the Bessel function of the first kind ($J$). In this paper, we will only need two transition amplitudes:
\bea[eq: transition amplitudes]
T^{+,+,-}=&T^{+,+,-}_{s}
\frac{\<\bar 1\bar 2\>^8}{\<\bar 1\bar 2\>^2\<\bar 2\bar 3\>^2\<\bar 3\bar 1\>^2}
\\
T^{+,+,+}=&T^{+,+,+}_{s}
\<\bar 1\bar 2\>^2\<\bar 2\bar 3\>^2\<\bar 3\bar 1\>^2
\eea 
which can be straightforwardly derived as is done in \cite{Raju:2012zs}.\footnote{$T^{+,+,\pm}_{s}$ are ``scalar'' parts of the transition amplitudes that depend only on the norms of the vectors; see \equref{eq: scalar part of transition amplitudes} for their explicit expressions.} The brackets refer to the products of spinor-helicities: we review our spinor helicity conventions in \S~{\ref{sec: spinor helicity formalism}}.

\subsection{Recursion relations in AdS}
\label{sec: recursion relations}
In this section, we will briefly review the recursion relations introduced in \cite{Raju:2012zr}, which generalizes the previous work of the author \cite{Raju:2010by,Raju:2011mp} in AdS and Risager's work in flat space \cite{Risager:2005vk}. 

Our starting point is the observation that we have
\be 
\sum\limits_{n\in\cN}\a_n\bep_n=0
\ee 
where $n$ parameterizes the external legs  and $\cN$ is a judiciously chosen set. This equality follows from the linear dependence of vectors for sufficiently high $\dim\cN$:\footnote{In low dimensions, such as $d=3$ (AdS$_4$), this is evidently true for four and higher point amplitudes. In higher dimensions, one can still proceed by decomposing the general polarization vector to a linear sum of special ones for which this equation is true, see \S 4.4 of \cite{Raju:2011mp}.} it allows the deformation of the external momenta while still preserving the momentum conservation:
\be 
\label{eq: deformation in vector notation}
\bk_n\rightarrow\bk_n(\w)=\left\{\begin{aligned}
	&\bk_n+\a_n\bep_n\w &n\in\cN 
	\\
	&\bk_n &\text{otherwise}
\end{aligned}\right.
\ee 

One can go ahead and use a contour deformation in the $\w-$plane to rewrite any tensor $T(\bk_1,\dots,\bk_m;\bep_1,\dots,\bep_m)$ in terms of its residues along with a possible boundary contribution. In the cause of gauge and gravity correlators, the Ward identities actually help us constraint the possible form of the boundary term; in the end, one is left with the following recursion relation:
\be 
\label{eq: recursion integration}
\hspace*{-.8em}
T(\bk_1,\dots,\bk_m;\bep_1,\dots,\bep_m)=\int\limits_0^\infty dp\left[\cB(0,p)+\sum\limits_\pi \cI_\pi(0,p)\right]
\hspace*{-.6em}
\ee 
where $\cB$ is a boundary term and $\pi$ is a bi-partitioning of the set $\{1,\dots,m\}$ into the sets $M_L$ and $M_R$ such that $M_{L,R}\cap\cN\ne\emptyset$. The piece $\cI_\pi(\w,p)$ can be reconstructed from lower-point \emph{transition amplitudes} as
\be 
\label{eq: integrand of recursion}
\hspace*{-1em}
\cI_\pi(\w,p)\propto \frac{-ip}{p^2+\left(\sum\limits_{n\in M_L}\bk_n(\w)\right)^2}\sum\limits_{\bep',\pm}T_\pi^2\frac{\w-\w_\pi^{\mp}(p)}{\w_\pi^{\pm}(p)-\w_\pi^{\mp}(p)}
\ee 
for
\begin{multline}
	\label{eq: T2 in vector form}
	T_\pi^2=T^*(\bk_{\pi_1}(\w_\pi^\pm),\dots,\bki(\w_\pi^\pm);\bep_{\pi_1},\dots,\bep')\\\x T^*(-\bki(\w_\pi^\pm),\dots,\bk_{\pi_m}(\w_\pi^\pm);\bep',\dots,\bep_{\pi_m})
\end{multline}
where $T^*$ is a \emph{transition amplitude} in the sense that it is same with $T$ upto the replacement of the bulk-to-boundary propagator of the momentum-$\bki$-particle with a \emph{normalizable mode}. The momentum $\bki$ is defined as
\be 
\label{eq: momentum of internal leg}
\bki(\w_\pi^\pm)=\sum\limits_{n\in M_L}\bk_n(\w_\pi^\pm)
\ee 
where $\w_\pi^\pm$ are solutions to the constraint \mbox{$\bki(\w_\pi^\pm)\cdot\bki(\w_\pi^\pm)=-p^2$}.

The algorithm of obtaining a higher point amplitude from a lower one is therefore:
\begin{enumerate}
	\item Turn lower point amplitude to a transition amplitude by replacing one of the bulk-to-boundary propagators with a normalizable mode
	\item Glue two such transition amplitudes, fixing the momentum of internal leg by \equref{eq: momentum of internal leg}
	\item Solve for $\w_\pi^\pm$: this is precisely what differentiates the curved space from the flat one as this relation is quadratic, unlike the linear relation of standard BCFW.
	\item Compute $\cI_\pi(\w,p)$ via \equref{eq: integrand of recursion}. We will focus on AdS$_4$, for which the relation will turn into an equality with a proportionality constant $1/4$ for the stress tensor correlator.\footnote{See eqn.~(4.12) of \cite{Raju:2012zs} for the more general case.}
	\item Sum over all possible permissable bi-partitions: this gives the integrand of higher point amplitude up to a constrained boundary piece.\footnote{The boundary piece is of the form	
		\be 
		\cB(\w,p)=\sum\limits_{m=0}^\infty a_m(p)\w^m +\text{piece-to-be-canceled}
		\ee 
		for rational functions $a_m(p)$, where the second piece cancels the divergent part of $\int \sum_\pi\cI_\pi dp$.
	}
	\item Do the integration in \equref{eq: recursion integration}. The integrand will be even in $p$ for all cases in this paper, therefore we can convert the integral into a residue integration where the contribution at the infinity is canceled precisely by the boundary piece $\cB$. 
\end{enumerate}

\section{Four point graviton amplitude in AdS$_4$}
\label{sec:main}
\subsection{Derivation of the general result}
We start by considering the polarization sum of the product of transitional amplitudes in the bi-partition $\pi=(12)(34)$. By using the identity \equref{eq: relation of scalar T} and some massaging, we get
\be 
\label{eq: helicity sum of transition amplitude product}
\sum\limits_{\bep}T^2=&\sum\limits_{i=\pm}T^{+,+,i}_{\bk_{1},\bk_2,\bki^\pm}T^{+,+,-i}_{\bk_{3},\bk_4,-\bki^\pm}
\\
=&
\left(1+\mathfrak{d}_3\right)
T^2_s\frac{\<\bar 1\bar 2\>^8\<\bar 3\bar 4\>^2\<\bar 4\lib\>^2\<\lib\bar 3\>^2}{\<\bar 1\bar 2\>^2\<\bar 2\lib\>^2\< \lib\bar 1\>^2}
\ee 
where $\li$ stands for the internal leg, and where $T_s^2$ and the operator $\mathfrak{d}_i$ are defined in the Appendix~\ref{sec: technical details}.

As shown in \equref{eq: T2 in vector form}, we actually need to deform the spinors as induced by the deformation of $\bk_i$ in \equref{eq: deformation in vector notation}{;}\footnote{We are considering all-line deformation, hence the set $\mathcal{N}$ is $\{1,2,3,4\}$.} in our case, this amounts to shifting the spinor $\lambda_i$ as
\begin{equation}
	\label{eq: deformation of spinors}
	(\lambda_m)_\a\;\rightarrow\; (\lambda_m)_\a+\w\;\beta_m\;\s^3_{\a\dot\a}(\lb_m)^{\dot\a}
\end{equation}
for $\beta$ given in \equref{eq: beta relations}, where external $\lb$'s remain the same. Thus, we only need to apply the deformation of the internal $\lib$,\footnote{We use the identities
	\bea 
	(\lb_{\li})_{\dot\a}=&-(\lb_2)_{\dot\a}-i\frac{k_1+k_2-\ki}{\<1^\pm2^\pm\>}\s^3_{\a\dot\a}\lambda_1^\a(\w^\pm)\;,
	\\
	(\lb_{\li})_{\dot\a}=&-(\lb_4)_{\dot\a}-i\frac{k_3+k_4+\ki}{\<3^\pm4^\pm\>}\s^3_{\a\dot\a}\lambda_3^\a(\w^\pm)
	\eea 
	derived in \cite{Raju:2012zs}.
} which leads to
\begin{multline}
	\label{eq: representation 2}
	\sum\limits_{\bep}T^2=
	\left(1+\mathfrak{d}_3\right)T^2_s\Bigg[
	\frac{\<\bar 1\bar 2\>^4\<\bar 3\bar 4\>[\bar 43^\pm]\<1^\pm2^\pm\>^2}{\<\bar 1\bar 2\>[\bar 21^\pm]\<3^\pm4^\pm\>^2}
	\\\x
	\frac{\<\bar 3\bar 4\>\<3^\pm4^\pm\>+2k_3(k_3+k_4+\ki)}{\<\bar 1\bar 2\>\<1^\pm2^\pm\>+2k_1(k_1+k_2-\ki)}\frac{k_3+k_4+\ki}{k_1+k_2-\ki}
	\Bigg]^2
\end{multline}
where we denote the deformed spinors by a superscript for brevity, i.e. $[1^\pm\bar 2]\equiv -(\s^3)^{\a\dot\beta}\left((\lambda_1)_\a+\w^\pm\s^3_{\a\dot\a}(\lb_1)^{\dot\a}\right)(\lb_2)_{\dot\beta}$ and so on. 

Alternatively, we could use momentum conservation to get rid of the explicit dependence on the internal leg $\lib$ first:\footnote{In spinors, $\bk_1+\bk_2+\bki=0$ reads as
	\be
	\label{eq: momentum conservation in spinors}
	\<a\li\>\<\lib\bar b\>=&i(k_1+k_2+\ki)[a\bar b]-\<a1\>\<\bar 1\bar b\>-\<a2\>\<\bar 2\bar b\>
	\ee
	which is used twice: first to rewrite ${\<\lib b\>}/{\<\lib a\>}$ as $\<c \li\>\<\lib a\>$, second to remove internal spinor completely.
} this leads to an expression with explicit $\lambda$'s (not just $\lb$'s as \equref{eq: helicity sum of transition amplitude product}), which deforms nontrivially to the final form
\begin{multline}
	\label{eq: representation 1}
	\sum\limits_{\bep}T^2=
	\left(1+\mathfrak{d}_3\right)
	T^2_s\<\bar 1\bar 2\>^6\<\bar 3\bar 4\>^2
	\bigg[
	\frac{\<1^\pm2^\pm\>^2\<\bar 2\bar 3\>\<\bar 4\bar 1\>}{(k_1+k_2+\ki)^2[1^\pm\bar 2][2^\pm\bar 1]}\\+i\frac{\<1^\pm2^\pm\>\<\bar 1\bar 4\>[1^\pm\bar 3]-\<1^\pm2^\pm\>\<\bar 2\bar 3\>[2^\pm\bar 4]}{(k_1+k_2+\ki)[1^\pm\bar 2][2^\pm\bar 1]}-\frac{[1^\pm\bar 3][2^\pm\bar 4]}{[1^\pm\bar 2][2^\pm\bar 1]}
	\bigg]^2
\end{multline}
We will make use of this alternative (yet equivalent) form below; nevertheless, let us proceed for now with the former expression. Inserting it into \equref{eq: integrand of recursion}, we get
\begin{multline}
	\cI_{(12)(34)}(0,p)=\left(1+\mathfrak{d}_3\right)\frac{ip}{4\left[p^2+\left(\bk_1+\bk_2\right)^2\right]}
	\\\x T^2_s\sum\limits_{\pm}\frac{\w_{12}^{\mp}(p)}{\w_{12}^{\pm}(p)-\w_{12}^{\mp}(p)}
	\Bigg[
	\frac{\<\bar 3\bar 4\>\<3^\pm4^\pm\>+2k_3(k_3+k_4+ip)}{\<\bar 1\bar 2\>\<1^\pm2^\pm\>+2k_1(k_1+k_2-ip)}
	\\\x
	\frac{k_3+k_4+ip}{k_1+k_2-ip}
	\frac{\<\bar 1\bar 2\>^4\<\bar 3\bar 4\>[\bar 43^\pm]\<1^\pm2^\pm\>^2}{\<\bar 1\bar 2\>[\bar 21^\pm]\<3^\pm4^\pm\>^2}
	\Bigg]^2
\end{multline}
where we used \equref{eq: action of D on w} to pull $\mathfrak{d}_3$ to the leftmost and set $\ki=ip$.\footnote{This is consistent both with normalizable mod having a \emph{negative} imaginary part required by the correct analytical continuation \cite{Raju:2012zr}, and with the constraint $\ki^2=-p^2$ in the recursion prescription.}

As we are interested in all plus helicity amplitude, different bi-partitions can be straightforwardly related to each other via $\mathfrak{d}$; in fact, by using \equref{eq: transformation of spinors} and
\equref{eq: action of D on w}, we can immediately write down $\sum_\pi\cI_\pi$ in the form $\sum_i\mathfrak{d}_i\cC$. All that is left is to insert this in \equref{eq: recursion integration}, which turns into a simple residue extraction as explained in the final item of the algorithm above; thus, putting everything together, we arrive at the final expression
\begin{multline}
	\label{eq: correlator final form}
	\hspace*{-1em}
	T^\text{MHV}(\bk_1,\dots,\bk_4)=\frac{i}{2^{11} k_1^2k_2^2k_3^2k_4^2}\left(\sum\limits_{i=0}^5\mathfrak{d}_i\right)\Big(
	\mathop{\mathrm{Res}}_{p=i(k_1+k_2)}
	+\mathop{\mathrm{Res}}_{p=i(k_3+k_4)}
	\\
	+\mathop{\mathrm{Res}}_{p=i\abs{\bk_1+\bk_2}}
	\Big)\Bigg\{
	\frac{\left(k_{12p^+}k_{12p^-}+2k_1k_2\right)\left(k_{34p^+}k_{34p^-}+2k_3k_4\right)}{\left[p^2+\left(\bk_1+\bk_2\right)^2\right]}
	\\\x
	\left(	\frac{k_{1^-2p}k_{12^-p}}{k_{12p^+}k_{12p^-}}\right)^2
	\sum\limits_{\pm}\frac{\w_{12}^{\mp}(p)}{\w_{12}^{\pm}(p)-\w_{12}^{\mp}(p)}
	\\
	\x\Bigg[
	\frac{\<\bar 3\bar 4\>\<3^\pm4^\pm\>+2k_3k_{34p^+}}{\<\bar 1\bar 2\>\<1^\pm2^\pm\>+2k_1k_{12p^-}}
	\frac{\<\bar 1\bar 2\>^4\<\bar 3\bar 4\>[\bar 43^\pm]\<1^\pm2^\pm\>^2}{\<\bar 1\bar 2\>[\bar 21^\pm]\<3^\pm4^\pm\>^2}
	\Bigg]^2
	\Bigg\}
\end{multline}
where we define
\be 
k_{a^{c_1}b^{c_2}p^{c_3}}=c_1k_a+c_2k_b+c_3ip
\ee 
for brevity, e.g. $k_{12p^-}=k_1+k_2-ip$.

One can compute the residues and insert the $\w^\pm$ computed in Appendix~\ref{sec: derivation of w}; to illustrate, let us consider the last residue. After a manageable calculation,\footnote{Although these computations are straightforward --albeit tediously long, we streamlined them like the rest of the calculations in this paper via the proprietary software \texttt{Mathematica} and the handy packages \texttt{xAct} \& \texttt{xTensor} \cite{garcia:2004xact,garcia:2004xtensor}.} one can show that the brackets at that residue satisfy
\begin{subequations}
	\label{eq: deformations to be replaced}
	\begin{align}
		\<1^-2^-\>=&\<12\>\;,\quad\<3^-4^-\>=\<34\>\\
		[\bar43^-]=&[\bar43]-\frac{\<\bar1\bar4\>[\bar21]}{\<\bar2\bar3\>}-\frac{\<\bar2\bar4\>[\bar12]}{\<\bar1\bar3\>}
		\\
		[\bar21^-]=&-\frac{[\bar12]\<\bar2\bar3\>\<\bar2\bar4\>}{\<\bar1\bar3\>\<\bar1\bar4\>}
	\end{align}
\end{subequations}
which yields the result
\begin{multline}
	\hspace*{-1em}
	T^\text{MHV}(\bk_1,\dots,\bk_4)\supset\frac{-\left(\sum\limits_{i=0}^5\mathfrak{d}_i\right)}{2^{8} k_1k_2k_3k_4}\Bigg\{
	\frac{\left(1-\frac{\cos(\theta_{12})}{2}\right)\left(1-\frac{\cos(\theta_{34})}{2}\right)}{\abs{\bk_1+\bk_2}}
	\\\x
	\Bigg[\cot^2\left(\frac{\theta_{12}}{2}\right)\frac{\<\bar 1\bar 2\>^3\<12\>^2\<\bar1\bar3\>\<\bar1\bar4\>\<\bar 3\bar 4\>}{[\bar12]\<\bar2\bar3\>\<\bar2\bar4\>\<34\>^2}
	\bigg([\bar43]-\frac{\<\bar1\bar4\>[\bar21]}{\<\bar2\bar3\>}
	\\
	-\frac{\<\bar2\bar4\>[\bar12]}{\<\bar1\bar3\>}\bigg)
	\frac{\<\bar 3\bar 4\>\<34\>+2k_3\left(k_3+k_4-\abs{\bk_3+\bk_4}\right)}{\<\bar 1\bar 2\>\<12\>+2k_1\left(k_1+k_2+\abs{\bk_1+\bk_2}\right)}
	\Bigg]^2
	\Bigg\}
\end{multline}
for $\cos(\theta_{ij})=\bk_i\cdot\bk_j/(k_ik_j)$. The other residues can be computed analogously: the full explicit result can be found in the attached Mathematica file.

\subsection{Center of mass frame}
\label{sec:center of mass frame}
The general result in \equref{eq: correlator final form} can be written in a simpler form by boosting to the center of mass frame, i.e. $\bk_1+\bk_2=\bk_3+\bk_4=0$.\footnote{We do not lose any generality here as this is an invertible transformation.} The identity $[i\bar j]=0$ for $\bk_i=-\bk_j$ ensures
\be 
\quad [\bar 12]=[\bar 21]=[\bar 34]=[\bar 43]=0
\ee 
in this frame, which simplify the overall computation. In particular, $\pi=(12)(34)$ partition gets the rather nice form
\be 
	\cI_{(12)(34)}(0,p)=\frac{-ip\left(1+\mathfrak{d}_3\right)}{4\left[p^2+\left(\bk_1+\bk_2\right)^2\right]}T_s^2\<\bar1\bar2\>^6\<\bar3\bar4\>^2\sum\limits_{n=-4}^4b_nc_n
\ee 
for $b_n$ defined in \equref{eq: definition of b coefficients}, where the only $\w^\pm$ dependence is carried by the following $c_n$ coefficients 
\be 
c_n=\left\{\begin{aligned}
	-\sum\limits_{i=0}^{n-2}(\w^+)^{n-i-1}(\w^-)^{i+1}\quad &\text{ for }n>1
	\\
	0\quad &\text{ for }n=1
	\\
	\sum\limits_{i=0}^{-n}(\w^+)^{n+i}(\w^-)^{-i}\quad &\text{ for }n<1
\end{aligned}\right.
\ee 

Despite not being in a polynomial form in $\w^\pm$, the other partitions are still simpler in this frame, e.g.
\begin{multline}
	\label{eq: 13 partition}
	\cI_{(13)(24)}(0,p)= \frac{ip}{4\left[p^2+\left(\bk_1+\bk_3\right)^2\right]}
	\\\x
	\sum\limits_{\pm}\frac{\w_{13}^{\mp}(p)}{\w_{13}^{\pm}(p)-\w_{13}^{\mp}(p)}\Bigg(
	\<\bar1\bar3\>^6\<\bar2\bar4\>^2\left(S_1(\w^\pm)\right)^2(\mathfrak{d}_1T_s^2)
	\\x
	+
	\<\bar2\bar4\>^6\<\bar1\bar3\>^2\left(S_2(\w^\pm)\right)^2(\mathfrak{d}_4T_s^2)
	\Bigg)
\end{multline}
where $S_{1,2}$ are known combinations of square and angle brackets, given in \equref{eq: details of S function}. 

\begin{figure*}
	\centering
	\begin{minipage}{.65\textwidth}
		\includegraphics[scale=.5]{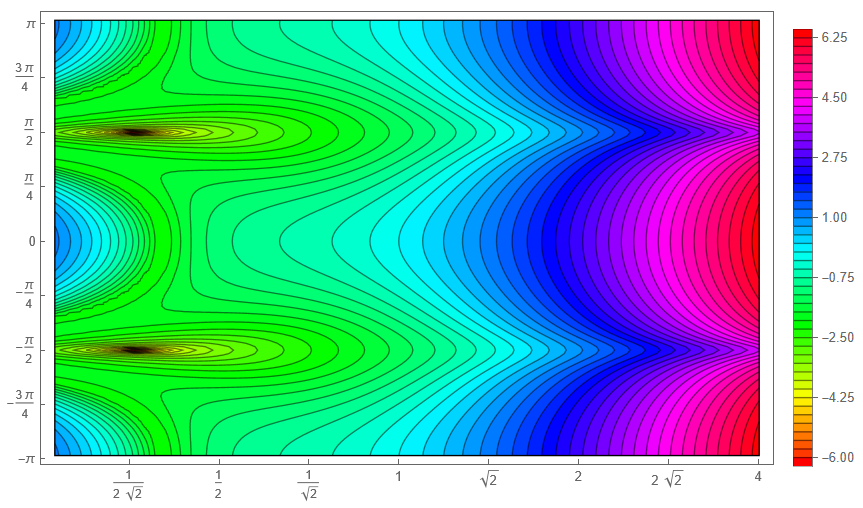}
	\end{minipage}
	\begin{minipage}{.32\textwidth}
		\caption{\label{figure}
			The contour plot of $\log\left(\abs{T(r,\theta)}/\abs{T(1,0)}\right)$ as a function of the parameters $r$ and $\theta$. We evaluated $T(r,\theta)$ at 31089 points in 200-digits precision and plotted with 50 contours. To avoid numerical problems with the spurious pole-zero cancellations, we slightly shifted $k_{2,3,4}$ to $\{1+10^{-45},r+10^{-50},r+10^{-55}\}$ instead of \{1,r,r\} respectively.
		}
	\end{minipage}
\end{figure*}
\subsection{Overall behavior of the amplitude}
In the previous section, we introduced the center of mass frame in which the amplitude takes a simpler form. We can further fix the configuration such that\footnote{By starting in a generic coordinate system, one can arrive at this configuration by \textbf{(1)} boosting to the center of mass frame, \textbf{(2)} scaling the zeroth coordinate so that $\bk^{(0)}_3-\bk^{(0)}_1=\abs{\bk}$, \textbf{(3)} shifting the zeroth coordinate so that $\bk^{(0)}_1=0$, \textbf{(4)} rotating the other coordinates so that $\bk^{(2)}_1=0$. As all of these transformations are invertible, the whole process amounts to fixing the symmetry without losing generality.}
\be 
\label{eq: frame}
\bk_1=-\bk_2=&\abs{\bk}(0,1,0)\;,\\ \bk_3=-\bk_4=&\abs{\bk}(1,\sqrt{1+r^2}\cos\theta,\sqrt{1+r^2}\sin\theta)
\ee 
where $\theta$ is the angle between the vectors $\bk_1$ and $\bk_3$,\footnote{Mathematically speaking, the actual \emph{angle} between the vectors $\bk_1$ and $\bk_3$ is $\arccos\left(\frac{\bk_1\cdot\bk_3}{k_1k_3}\right)=\arccos\left(\sqrt{1+\frac{1}{r^2}}\cos\theta\right)$; however, we will refer to $\theta$ as the angle as it measures the \emph{collinearity} of $\bk_1$ and $\bk_3$ in the spatial coordinates.
} and $r$ is the ratio of the norms $k_3$ and $k_1$. In this frame, we have
\be 
T(\bk_1,\dots,\bk_4)\evaluated_{\text{in frame \eqref{eq: frame}}}=\abs{\bk}^3T(r,\theta)
\ee 
where we can explicitly compute $T(r,\theta)$; for instance, the piece in front of $\mathfrak{d}_0$ becomes
\begin{multline}
	T(r,\theta)\supset -\frac{2 e^{-4 i \theta } (r+1) \left(r^2+1\right)^2 (2 r^4+2 r^3-7 r^2+2 r+2) }{(r-1)^2 r^2}
	\\\x\left(\left(r^2+1\right) \cos (2 \theta )+1\right)
\end{multline}
However, the other pieces are more complicated and it is in fact not sensible to consider individual pieces as they develop spurious poles in this frame.\footnote{We physically expect (and numerically verify) that the full amplitude does not have poles at the locations $\bk_1+\bk_2$ and $\bk_3+\bk_4$; hence we know the singularities in the individual contributions to be spurious.} Therefore, we consider the full amplitude instead, and extract its overall behavior with respect to its arguments numerically, which we present in Figure~(\ref{figure}). We provide the corresponding data in the attached Mathematica file.

\section{Conclusions}
\label{sec:conclusion}
In this paper, we symbolically computed the four point graviton amplitude in AdS for all plus external helicities. We did this by using the AdS on-shell recursion relations, which have seen limited attention and have not been used in explicit full computations since their introduction more than a decade ago.\footnote{At the same time of his introduction of this technology in \cite{Raju:2012zr}, the author publishes another paper in which he actually carries out explicit computation of four point graviton amplitude of $\{+,-,+,-\}$ helicities \cite{Raju:2012zs}. However, that computation does not fairly represent the technical difficulty of this technology, because most of the partitions for (and only for) this set of helicities have deformations free of square roots (main complication of quadratic solution for deformation).
} As we mentioned in the introduction, on-shell recursion relations in curved spaces are far more complicated than their flat space counterparts, and this work is intended to serve the goal of being a playground to test the strengths and weaknesses of BCFW on AdS. 

We found out that RBCFW is rather efficient at providing compact \emph{implicit} results, as we did with \equref{eq: correlator final form}. Indeed, we were able to package the formidable four point graviton amplitude in AdS to mere four lines without the use of any four point interaction vertices. However, to get the full \emph{explicit} result, one needs to extract out the deformations (see \equref{eq: deformations to be replaced}) and compute the residues, which are straightforward albeit tedious computations. One still needs to take these residues in the usual diagrammatic approach, since Witten diagrams also have analogous bulk integrals; nevertheless, the deformation itself is rather lengthy in the general case as we detail in \S~\ref{sec: derivation of w}. However, all of the necessary steps are highly algorithmic, which takes us to our next point.

We have observed that the rigid algorithmic approach in RBCFW (as loosely listed at the end of \S~\ref{sec:preliminaries}) makes it rather straightforward to implement the calculations into a symbolic computation software. Indeed, we wrote a short script in the propriety program \texttt{Mathematica} to carry out the computations: this allowed us to obtain the \emph{full, explicit} result, which we provide in the attached file. The downside though is that the result is \emph{not} yet in the shortest possible form,\footnote{It is relatively hard to teach Mathematica to make use of all Schouten identities and momentum conservations in an intelligent and time-efficient way; perhaps the machine learning can be utilized for such optimization tasks in future.} which makes it improbable to extract a physical intuition.

Besides computing the amplitude, we also commented with explicit results on how certain configurations simplify the calculations; nevertheless, they are still too lengthy to be performed by pen and paper. We believe that certain limits should make them more manageable, and we plan to explore this in our next project. It is of course plausible as well to compute higher point amplitudes and see how the complexity of explicit results change with that. Furthermore, one could introduce machine learning or more general AI tools to explore various ways to train the software to simplify the lengthy results into forms from which clearer physical intuitions can be extracted. Additionally, it is entirely possible that using the right variable rather than the three dimensional spinor helicity might be the best way to move forward: we believe that more attention needs to go into the investigation of the right variables to approach this problem. 

These points aside, we do see some simplicity. Much like in Raju's work, perhaps the most physical pole coming from the propagator in the four point diagrams yields the simplest result, leading to cancellation of all the messy terms. However, we still need to understand how best to simplify the expressions coming from the poles of the lower point amplitudes. In summary, on-shell recursion relations in AdS can really be useful to  get the full explicit results in a reliable and fast fashion, \emph{without} caring for the redundancies in the result.\footnote{As noted before, our current implementation does not make much use of various identities.} They are also quite useful to compute numerical results, which are valuable data to compare with experiments. These advantages become radically more significant in the calculation of higher point amplitudes since the approach is sufficiently robust and efficiently recursive. Additionally, they could be a useful device to check double copy relations for higher point amplitudes in (A)dS \cite{Lee:2022fgr, Li:2022tby, Drummond:2022dxd, Herderschee:2022ntr, Sivaramakrishnan:2021srm, Diwakar:2021juk, Zhou:2021gnu, Albayrak:2020fyp, Armstrong:2020woi}. 

Before closing, we will share a poignant anecdote from DeWitt.\footnote{We were inspired to mention this historical allegory after coming across a similar and uplifting tale in a related paper, where we found a quote from Parke and Taylor on page 39 \cite{Bonifacio:2022vwa}.} In 1968, Bryce DeWitt impressively computed the four-point scattering amplitudes for gravitons \cite{DeWitt:1967uc}. In the paper, DeWitt writes a sagacious commentary on the subject that remains striking to this day. He says, ``it is a pity that nature displays such indifference to so intriguing and beautiful a subject [graviton scattering], for the calculations themselves are of considerable intrinsic interest.'' DeWitt later says ``there must be an easier way'' due to the significant amount of cancellation between terms. Despite the fact that this quote has been partially buried in the passage of time, posterity has ultimately vindicated his belief, about the simplicity of these structures, to be correct.
Admittedly, in modern times, we have at our disposal an exceedingly straightforward formula for graviton scattering amplitudes involving $n$ points \cite{Hodges:2012ym}. It is eminently plausible that we stand at the threshold of generalizing these breakthroughs to encompass curved spaces such as (A)dS.

\begin{acknowledgments}
We would like to thank Chandramouli Chowdhury, Jinwei Chu, Nikos Dokmetzoglou, Austin Joyce, Hayden Lee, and David Meltzer for discussions. SA is supported by a VIDI grant of the Netherlands Organisation for Scientic Research (NWO) that is funded by the Dutch Ministry of Education, Culture and Science (OCW).
\end{acknowledgments}

\appendix 
\section{}
\setcounter{equation}{0}
\renewcommand\theequation{A.\arabic{equation}}

\subsection{Spinor Helicity Formalism}
\label{sec: spinor helicity formalism}
A vector $\bm{v}$ in the flat $3d$ boundary of the AdS$_4$ can be turned into a null $4d$ vector by attaching its norm as a fourth component, i.e. $(\vec{v},i\abs{\vec{v}})$. One can then convert this vector into a pair of spinors $\lambda$ and $\lb$ by the relation
\be 
\lambda_\a[\bm{v}]\lb_{\dot\a}[\bm{v}]=\bm{v}_0\s^0_{\a\dot\a}+\bm{v}_1\s^1_{\a\dot\a}+\bm{v}_2\s^2_{\a\dot\a}+i\sqrt{\bm{v}_\mu \bm{v}^\mu}\s^3_{\a\dot\a}
\ee 
where we define $\s$-matrices as
\be 
\s_{\a\dot\a}^{0,\dots,3}=\left\{
\begin{pmatrix}
	1&0\\0&1
\end{pmatrix},
\begin{pmatrix}
	0&1\\1&0
\end{pmatrix},
\begin{pmatrix}
	0&-i\\i&0
\end{pmatrix},\begin{pmatrix}
	1&0\\0&-1
\end{pmatrix}
\right\}
\ee 
In the rest of the paper we use the shorthand notation $(\lambda_i)_{\a}$ instead of $\lambda_\a[\bk_i]$ for clarity.

Unlike its flat space counterpart, the relevant symmetry group here is $\SO(2,1)$, not $\SO(3,1)$: that means the relevant isogeny is \emph{not} $\SL(2,\C)\rightarrow\SO(3,1)$, but $\SL(2,\R)\rightarrow\SO(2,1)$. In practice, this amounts to the existence of an invariant combination of $\lambda$ and $\lb$ which we denote as\footnote{Despite the presence of the relative minus sign in our definition of $[i,\bar j]$ compared to \cite{Raju:2012zs}, our conventions are same: the author there chooses $(\s^3)^{\a\dot\a}=(\s^3)_{\a\dot\a}=\mathrm{diag}\left(1,-1\right)$, whereas we choose $(\s^3)_{\a\dot\a}=\mathrm{diag}\left(1,-1\right)$ and $(\s^3)^{\a\dot\a}=\e^{\a\beta}\e^{\dot\a\dot\beta}(\s^3)_{\beta\dot\beta}$. In summary, all of our results are consistent with his conventions when written in terms of brackets.}
\be 
[i\bar j]=-(\s^3)^{\a\dot\beta}(\lambda_i)_{\a}(\lb)_{\dot\beta}
\ee 
Of course, we still have the familar invariants of the flat space scattering amplitudes as well:
\be 
\<ij\>=\e^{\a\beta}(\lambda_i)_\a(\lambda_j)_\beta\;,\quad 
\<\bar i\bar j\>=\e^{\dot\a\dot\beta}(\lb_i)_{\dot\a}(\lb_j)_{\dot\beta}
\ee 
for $\e-$tensor
\be 
\e^{\a\beta}=\e^{\dot\a\dot\beta}=\e_{\a\beta}=\e_{\dot\a\dot\beta}=\begin{pmatrix}
	0&1\\-1&0
\end{pmatrix}
\ee 
We can raise and lower the indices with the $\e$ by contracting from left and right respectively:
\be 
\lambda^\a=\e^{\a\beta}\lambda_\beta\quad,\quad \lambda_\a=\lambda^{\beta}\e_{\beta\a}
\ee 
and analogously for the dotted indices.

One can work out relations between products of $\s^3$ and $\e$ via their explicit representations:
\be 
\e_{\a\beta}\e_{\rho\de}+\e_{\rho\a}\e_{\beta\de}+\e_{\de\a}\e_{\rho\beta}=&0\\
\e_{\a\beta}\e_{\dot\a\dot\beta}+\s^3_{\a\dot\a}\s^3_{\beta\dot\beta}-\s^3_{\a\dot\beta}\s^3_{\a\dot\beta}=&0\\
\dots&
\ee 

Contractions of these equations with arbitrary spinors produce generalizations of the Schouten identities in the scattering amplitudes literature; for instance,
\be 
\label{eq: schouten 1}
\<ij\>\<\bar i\bar j\>=[i\bar j][j\bar i]-[i\bar i][j\bar j]
\ee

\subsection{Technical Details}
\label{sec: technical details}
We define the scalar part of the three point graviton transition amplitudes in \equref{eq: transition amplitudes} as
\bea[eq: scalar part of transition amplitudes]
T^{+,+,-}_{s}=&\sqrt{\frac{2}{\pi}}\frac{\left(k_1^2+k_2^2+4k_1k_2-k_3^2\right)(-k_1+k_2+k_3)^2}{32k_1^2k_2^2\sqrt{-ik_3}\left(k_1^2+k_2^2+2k_1k_2-k_3^2\right)^2}
\nn\\&\x(k_1-k_2+k_3)^2(k_1+k_2-k_3)^2
\\
T^{+,+,+}_{s}=&\sqrt{\frac{2}{\pi}}\frac{\left(k_1^2+k_2^2+4k_1k_2-k_3^2\right)\left(k_1+k_2+k_3\right)^2}{32k_1^2k_2^2\sqrt{-ik_3}\left(k_1^2+k_2^2+2k_1k_2-k_3^2\right)^2}
\eea
where $k_3$ is on a different footing than $k_{1,2}$ as it corresponds to a normalizable mode. Note that, these factors satisfy the relation $\sqrt{ik_3}T^{+,+,\pm}_{s;k_1,k_2,-\ki}=\sqrt{-ik_3}T^{+,+,\pm}_{s;-k_1,-k_2,\ki}$.
When viewed as a mere three point transition amplitude, $\ki$ corresponds to the norm associated with the momentum of the normalizable mode, which is only defined for timelike vectors, i.e. $\ki^2<0$. $\ki$ needs to be imaginary also in the recursion relation of higher point functions, as we impose $\ki^2=-p^2$ for $p\in\R$. Thus, we end up with the relation 
\be 
\label{eq: relation of scalar T}
T^{+,+,\pm}_{s;k_1,k_2,-\ki}=-i\;T^{+,+,\pm}_{s;-k_1,-k_2,\ki}
\ee 
This identity is rather useful, as it allows to simplify the helicity sum in \equref{eq: helicity sum of transition amplitude product}. The term $T_s^2$ there is defined as
\begin{multline}
	\label{eq: definition of T^2_s}
	T^2_s=
	\frac{\left(k_1^2+k_2^2+4k_1k_2-\ki^2\right)\left(k_3^2+k_4^2+4k_3k_4-\ki^2\right)}{2^9\pi k_1^2k_2^2k_3^2k_4^2\ki}
	\\\x\frac{(-k_1+k_2+\ki)^2(k_1-k_2+\ki)^2}{\left(k_1+k_2+\ki\right)^2\left(k_3+k_4+\ki\right)^2}
\end{multline}

For convenience with the later definitions, let us define the following operator:
\be 
\cD^{i,j,k,l}_{a,b,c,d}\cO=\cO\evaluated_{
	\substack{
		\lb_1\rightarrow\lb_i,\lb_2\rightarrow\lb_j,\lb_3\rightarrow\lb_k,\lb_4\rightarrow\lb_l
		\\
		\lambda_1\rightarrow\lambda_a,\lambda_2\rightarrow\lambda_b,\lambda_3\rightarrow\lambda_c,\lambda_4\rightarrow\lambda_d		
}}
\ee 
A minus sign in the indices indicates a sign change in the replacement, i.e. $\cD^{-2,1,3,4}_{1,2,3,4}$ would replace $\lb_1$ with $-\lb_2$. In this paper, we only need the following special instances of this operator:
\be 
\label{eq: definition of lowercase d}
\mathfrak{d}_1=\cD^{1324}_{1324}\quad\mathfrak{d}_2=\cD^{1423}_{1423}\quad\mathfrak{d}_3=\cD_{-3,-4,-1,-2}^{3,4,1,2,}\\\mathfrak{d}_4=\cD^{2413}_{-2-4-1-3}\quad\mathfrak{d}_5=\cD_{-2-3-1-4}^{2314}\qquad
\ee 
where we also define $\mathfrak{d}_0=\cD^{1234}_{1234}=1$ for consistency. Note that these operators also act on the norms of vectors as they are related to the square brackets, i.e.
\be 
k_j=-\frac{i}{2}[j\bar j]
\ee 
The action of $\cD$ on everything else conforms to the naive expectations; for instance,
\be 
\label{eq: transformation of beta}
\cD^{ijkl}_{abcd}\;:\quad \beta_1\rightarrow\beta_i\;,\dots,\;\beta_4\rightarrow\beta_l
\ee 
which also implies
\be 
\label{eq: transformation of spinors}
\cD^{ijkl}_{\pm i bcd}\lambda_1(\w)=\pm\lambda_i(\cD^{ijkl}_{\pm i bcd}\w)
\ee 
and analogously for other $\lambda_i$. This also leads to the neat relation
\be 
\label{eq: action of D on w}
\cD^{ijkl}_{abcd}\w_{12}^\pm=\w_{ij}^\pm\quad\text{ if }\quad (i,j,k,l)=\pm(a,b,c,d)
\ee 
where all relevant $\mathfrak{d}_i$ obey the conditional.\footnote{One can show this by comparing the action of the operator $\cD^{ijkl}_{abcd}$ in the partition $\pi=(12)(34)$ of \equref{eq: definition of wpm}, with the partition $\pi=(ij)(..)$ of the same equation.
} Note that similar identities apply for other $\w_{ij}^\pm$, e.g. $\cD_{1,3,2,4}^{1,3,2,4}\w_{34}^+=\w_{24}^+=\w_{13}^+$.\footnote{The second equality follows from the trivial symmetries of $\w_{ij}^\pm$, i.e. 
	\be 
	\w_{ij}^\pm=\w^{\pm}_{ji}\;,\quad	\w_{ij}^\pm=\w^{\pm}_{i'j'}\text{ for }i\ne j\ne i'\ne j'\in\{1,2,3,4\}
	\ee } This allows one to compute one piece of the stress tensor correlator and apply $\cD$ operators at the end to get all contributions, as explicitly shown in \equref{eq: correlator final form}.

\subsection{Derivation of $\w^\pm$}
\label{sec: derivation of w}
The defining equation of $\w^\pm$  is \mbox{$\bki(\w^\pm)\cdot\bki(\w^\pm)=-p^2$}; the momentum conservation then turns this into the relation\footnote{We are also making use of the identity
	\be 
	\bk\cdot\bm{q}=-\half\<kq\>\<\bar k\bar q\>-\frac{1}{4}[k\bar k][q\bar q]
	\ee 
}
\be 
\label{eq: definition of wpm}
\<\pi_1(\w_\pi^\pm),\pi_2(\w_\pi^\pm)\>\<\bar \pi_1(\w_\pi^\pm),\bar \pi_2(\w_\pi^\pm)\>=(k_{\pi_1}+k_{\pi_2})^2+p^2
\ee 
for the deformation given in \equref{eq: deformation of spinors}. The $\beta-$coefficients there are fixed upto an overall scaling with the relations
\be
\label{eq: beta relations}
\hspace*{-1em}
\begin{aligned}
	\frac{\beta_2}{\beta_1}=-\frac{\<\bar 1 \bar 4\>\<\bar1 \bar 3\>}{\<\bar 2\bar4\>\<\bar2\bar3\>}\;,\; 
	\frac{\beta_3}{\beta_1}=-\frac{\<\bar1\bar4\>\<\bar1\bar2\>}{\<\bar3\bar4\>\<\bar3\bar2\>}
	\\ 
	\frac{\beta_4}{\beta_1}=-\frac{\<\bar1\bar2\>\<\bar1\bar3\>}{\<\bar4\bar2\>\<\bar4\bar3\>} 
	\;,\;
	\frac{\beta_4}{\beta_3}=-\frac{\<\bar3\bar2\>\<\bar3\bar1\>}{\<\bar4\bar2\>\<\bar4\bar1\>}
\end{aligned}
\hspace*{-1em}
\ee 

One can go ahead and solve the equation to get the explicit result for $\w^\pm$ as a function of $p$; however, we only need $\w^\pm(p^*)$ for the location of the poles $p^*$. Below, we list $\w_{12}^\pm$ at all poles, from which other $\w_{ij}^\pm$ can be computed.
\bea 
{}&\w^\pm_{12}\evaluated_{p=i(k_1+k_2)}=\frac{-\beta_1[2\bar 1]+\beta_2[1\bar2]}{2\beta_1\beta_2\<\bar1\bar2\>}
\nn\\
{}&\qquad\quad\mp\frac{\sqrt{
		\left(
		-\beta_1[2\bar1]+\beta_2[1\bar2]
		\right)^2+4\beta_1\beta_2\<12\>\<\bar1\bar2\>
}}{2\beta_1\beta_2\<\bar1\bar2\>}
\\
{}&\w^\pm_{12}\evaluated_{p=i(k_3+k_4)}=-\mathfrak{d}_3\left(\w^\mp_{12}\evaluated_{p=i(k_1+k_2)}\right)
\\
{}&\w^+_{12}\evaluated_{p=i\abs{\bk_1+\bk_2}}=0
\\
{}&\w^-_{12}\evaluated_{p=i\abs{\bk_1+\bk_2}}=
\frac{-\beta_1[\bar 12]+\beta_2[\bar 2 1]}{\beta_1\beta_2\<\bar 1\bar 2\>}
\eea 
These can be compared to the similar computations in \cite{Raju:2012zs}.\footnote{Note that the author uses \emph{hatted} spinors which we do not introduce in this paper. The translation between brackets of such spinors and others can be made via the identities $\<\bar i\hat j\>=[j\bar i]$ and $\<\bar{\hat i}\bar{\hat j}\>=-\<\bar i\bar j\>$.}

\subsection{Coefficients in the center of mass frame}
In the frame detailed in Section~\ref{sec:center of mass frame}, \equref{eq: representation 1} becomes
\be
\sum\limits_{\bep}T^2=
\left(1+\mathfrak{d}_3\right)
T^2_s\<\bar 1\bar 2\>^6\<\bar 3\bar 4\>^2
\left[\sum\limits_{n=-2}^2a_n\w^n
\right]^2
\ee 
for
\bea 
a_2=&-\frac{\left\langle \bar{1}\bar{3}\right\rangle  \left\langle \bar{1}\bar{4}\right\rangle ^2}{\left(2 k_1+i p\right){}^2 \left\langle \bar{2}\bar{4}\right\rangle }
\\
a_1=&\frac{2  \left\langle \bar{1}\bar{3}\right\rangle  \left\langle \bar{1}\bar{4}\right\rangle }{\left(-p+2 i k_1\right) \left\langle \bar{1}\bar{2}\right\rangle }
\\
a_0=&\frac{-2 \langle 12\rangle  \left\langle \bar{1}\bar{4}\right\rangle  \left\langle \bar{2}\bar{3}\right\rangle +i \left(2 k_1+i p\right) \left(\left\langle \bar{1}\bar{4}\right\rangle  \left| 1\bar{3}\right| -\left\langle \bar{2}\bar{3}\right\rangle  \left| 2\bar{4}\right| \right)}{\left(2 k_1+i p\right)^2\left\langle \bar{1}\bar{2}\right\rangle}
\nn\\&+\frac{\left\langle \bar{1}\bar{3}\right\rangle  \left\langle \bar{2}\bar{4}\right\rangle }{\left\langle \bar{1}\bar{2}\right\rangle ^2}
\\
a_{-1}=&-\frac{\left(2 k_1+i p\right) \left(\left\langle \bar{1}\bar{4}\right\rangle  \left| 1\bar{3}\right| -\left\langle \bar{2}\bar{3}\right\rangle  \left| 2\bar{4}\right| \right)\left\langle \bar{2}\bar{4}\right\rangle }{\left(2 k_1+i p\right) \left\langle \bar{1}\bar{2}\right\rangle ^2 \left\langle \bar{1}\bar{4}\right\rangle }
\nn\\
&-2 i\frac{ \langle 12\rangle  \left\langle \bar{1}\bar{4}\right\rangle  \left\langle \bar{2}\bar{3}\right\rangle \left\langle \bar{2}\bar{4}\right\rangle }{\left(2 k_1+i p\right) \left\langle \bar{1}\bar{2}\right\rangle ^2 \left\langle \bar{1}\bar{4}\right\rangle }
\\
a_{-2}=&
-\frac{\left\langle \bar{2}\bar{3}\right\rangle  \left\langle \bar{2}\bar{4}\right\rangle  \left(\langle 12\rangle  \left\langle \bar{2}\bar{3}\right\rangle +\left(p-2 i k_1\right) \left| 1\bar{3}\right| \right)}{\left(2 k_1+i p\right){}^2 \left\langle \bar{1}\bar{2}\right\rangle ^2 \left\langle \bar{1}\bar{3}\right\rangle  \left\langle \bar{1}\bar{4}\right\rangle }
\nn\\ 
&\x\left(\langle 12\rangle  \left\langle \bar{1}\bar{4}\right\rangle +\left(-p+2 i k_1\right) \left| 2\bar{4}\right| \right)
\eea 
We can then go ahead and define the coefficients $b_n$ by the relation
\be 
\label{eq: definition of b coefficients}
\sum\limits_{n=-4}^4b_n\w^n=\left[\sum\limits_{n=-2}^2a_n\w^n
\right]^2
\ee 
Finally, we define the functions $S_{1,2}(\w)$ referred in \equref{eq: 13 partition} as follows:
\small 
\begin{widetext}
	\bea[eq: details of S function]
	\hspace*{-2em}
	S_1(\w)=&\frac{\left(\omega  \left\langle \bar{1}\bar{2}\right\rangle  \left(-\left\langle \bar{1}\bar{4}\right\rangle  \left| 1\bar{3}\right| +i \left(k_1+k_3+i p\right) \left\langle \bar{3}\bar{4}\right\rangle \right)+\omega ^2 \left\langle \bar{1}\bar{2}\right\rangle  \left\langle \bar{1}\bar{3}\right\rangle  \left\langle \bar{1}\bar{4}\right\rangle -\left\langle \bar{2}\bar{3}\right\rangle  \left\langle \bar{3}\bar{4}\right\rangle  \left(\langle 13\rangle -\omega  \left| 3\bar{1}\right| \right)\right)}{\left(k_1+k_3+i p\right){}^2 \left\langle \bar{3}\bar{4}\right\rangle  \left(\omega  \left\langle \bar{1}\bar{3}\right\rangle -\left| 1\bar{3}\right| \right) \left(-\omega  \left\langle \bar{1}\bar{2}\right\rangle  \left\langle \bar{1}\bar{3}\right\rangle  \left\langle \bar{1}\bar{4}\right\rangle -\left\langle \bar{2}\bar{3}\right\rangle  \left\langle \bar{3}\bar{4}\right\rangle  \left| 3\bar{1}\right| \right)}
	\nn\\&\x \left(\omega  \left\langle \bar{1}\bar{2}\right\rangle  \left\langle \bar{1}\bar{4}\right\rangle  \left(-\left\langle \bar{1}\bar{4}\right\rangle  \left| 1\bar{3}\right| +i \left(k_1+k_3+i p\right) \left\langle \bar{3}\bar{4}\right\rangle \right)+\omega ^2 \left\langle \bar{1}\bar{2}\right\rangle  \left\langle \bar{1}\bar{3}\right\rangle  \left\langle \bar{1}\bar{4}\right\rangle ^2-\left\langle \bar{2}\bar{3}\right\rangle  \left\langle \bar{3}\bar{4}\right\rangle  \left\langle \bar{1}\bar{4}\right\rangle  \left(\langle 13\rangle -\omega  \left| 3\bar{1}\right| \right)\right)
	\\
	S_2=&\frac{\left(\omega  \left(-\left\langle \bar{1}\bar{2}\right\rangle \right) \left(\left\langle \bar{2}\bar{3}\right\rangle  \left| 2\bar{4}\right| -i \left(-k_2-k_4+i p\right) \left\langle \bar{3}\bar{4}\right\rangle \right)-\left\langle \bar{3}\bar{4}\right\rangle  \left(-\left\langle \bar{1}\bar{4}\right\rangle  \left(\omega  \left| 4\bar{2}\right| +\langle 24\rangle \right)-\left(p-i \left(-k_2-k_4\right)\right) \left| 2\bar{1}\right| \right)-\omega ^2 \left\langle \bar{1}\bar{2}\right\rangle  \left\langle \bar{2}\bar{3}\right\rangle  \left\langle \bar{2}\bar{4}\right\rangle \right)}{\left(-k_2-k_4+i p\right){}^2 \left\langle \bar{3}\bar{4}\right\rangle  \left(\omega  \left\langle \bar{2}\bar{4}\right\rangle +\left| 2\bar{4}\right| \right) \left(\omega  \left\langle \bar{1}\bar{2}\right\rangle  \left\langle \bar{2}\bar{3}\right\rangle  \left\langle \bar{2}\bar{4}\right\rangle -\left\langle \bar{1}\bar{4}\right\rangle  \left\langle \bar{3}\bar{4}\right\rangle  \left| 4\bar{2}\right| \right)}
	\nn\\&\x\left(-\omega  \left\langle \bar{1}\bar{2}\right\rangle  \left\langle \bar{2}\bar{3}\right\rangle  \left(\left\langle \bar{2}\bar{3}\right\rangle  \left| 2\bar{4}\right| -i \left(-k_2-k_4+i p\right) \left\langle \bar{3}\bar{4}\right\rangle \right)+\left\langle \bar{1}\bar{4}\right\rangle  \left\langle \bar{3}\bar{4}\right\rangle  \left(\left\langle \bar{2}\bar{3}\right\rangle  \left(\omega  \left| 4\bar{2}\right| +\langle 24\rangle \right)-i \left(-k_2-k_4+i p\right) \left| 4\bar{3}\right| \right)-\omega ^2 \left\langle \bar{1}\bar{2}\right\rangle  \left\langle \bar{2}\bar{4}\right\rangle  \left\langle \bar{2}\bar{3}\right\rangle ^2\right)
	\eea
\end{widetext}
\normalsize

\bibliography{collectiveReferenceLibrary.bib}
\bibliographystyle{utphysShorter}
\end{document}